\documentclass[preprint,12pt,3p]{elsarticle}
\usepackage{amssymb}
\usepackage{mathtools}
\usepackage{amsmath}
\usepackage{graphicx}
\usepackage{booktabs}
\usepackage{url}
\usepackage{soul}
\usepackage{xcolor}

\journal{Renewable Energy}
\begin{document}

\let\WriteBookmarks\relax
\def\floatpagepagefraction{1}
\def\textpagefraction{.001}
\begin{frontmatter}

\title{Internal hydro- and wind portfolio optimisation in real-time market operations} 

\author[label1,label2]{Hans Ole Riddervold\corref{cor1}}
\address[label1]{NTNU}
\address[label2]{Norsk Hydro\fnref{label4}}
\ead{hans.o.riddervold@ntnu.no}
\ead[url]{https://www.ntnu.no/ansatte/hans.o.riddervold}

\author[label5]{Ellen Krohn Aasgård}
\address[label5]{SINTEF}
\ead{Ellen.Aasgard@sintef.no}

\author[label2]{Lisa Haukaas}
\ead{Lisa.Haukaas@hydro.com}

\author[label1]{Magnus Korpås}
\ead{magnus.korpas@ntnu.no}

\cortext[cor1]{Corresponding author}

\begin{abstract}
In this paper aspects related to handling of intraday imbalances for hydro and wind power are addressed. The definition of imbalance cost is established and used to describe the potential benefits of shifting from plant-specific schedules to a common load requirement for wind and hydropower units in the same price area. The Nordpool intraday pay-as-bid market has been the basis for evaluation of imbalances, and some main characteristics for this market has been described. 
We consider how internal handling of complementary imbalances within the same river system with high inflow uncertainty and constrained reservoirs can reduce volatility in short-term marginal cost and risk compared to trading in the intraday market.
We have also shown the the imbalance cost for a power producer with both wind and hydropower assets can be reduced by internal balancing in combination with sales and purchase in a pay-as-bid intraday market. 
\end{abstract}

\begin{keyword}
Wind power \sep Reservoir hydro \sep Marginal cost \sep Balancing \sep Intraday \sep Pay-as-bid

\end{keyword}

\end{frontmatter}















\section{Introduction}

With the increasing penetration of wind in the European market, power producers are in a larger extent managing combined portfolios of wind- and hydropower. In this context it is interesting to evaluate the value of internal coordination for planning and balancing of combined portfolios owned or operated by the the same company.   

Basic economic theory suggest that as long as all market actors bid production at marginal cost, the optimal balance
of consumption and production will be established in the market. This further implies that there should be no or limited need for internal balancing of production within a company, and that all power plants 
should bid their marginal cost of production individually. The concept can be exemplified with a company owning two plants. If one plant ends up with an expected imbalance caused by wind power production below the original prediction, there is no need to compensate by changing production in another plant as long as the expected imbalance volume can be bought or sold at lower cost. If it by chance should be the internal plant who provides the least costly solution, this would any how be found as long as this production is provided to the market place. 

However, there are several challenges that might hinder a company managing several production units to find the optimal solution by only interacting with an external market.  These challenges can be:

\begin{itemize}
    \item Limited liquidity and large spreads discouraging market participation
    \item Interacting with a pay-as-bid versus a marginal cost market which has been a historic reference for many power producers
    \item Real-time pricing and validity for hydropower based short term marginal cost
    \item Establishing marginal- / opportunity cost for wind as basis for pricing in the intraday market
    \item Time for placement of bids with continuously shifting market prices, volumes  
    , inflow and wind.  
    \item Access to non-public information through observations and operational data within a portfolio
    \item Emergence of autonomous trading systems and algorithmic trading 
    \item Trading- and imbalance fees associated with market interaction 
\end{itemize}

The development and application of the virtual power plant (VPP) concept during the last decade show that there has been a demand in the market for actors willing to take on a role as coordinator or aggregator between distributed energy production and the market. Several articles have addressed the topic of modelling VPP's \cite{NAVAL202057, ZAPATARIVEROS2015408}, or related concepts of bidding though an external agent \cite{7276961}. 

In section 2, a description of the market conditions forming the basis for this analysis are described together with some relevant market aspects that might influence the choice of whether to handle imbalance internally or in the market. Further,  operational challenges from the perspective of a power producer managing a combined portfolio of wind- and hydropower are addressed. In section 3, a method is proposed for evaluating the benefits of internal balancing in combination with the intraday market. This method is applied on a case study in section 4 before drawing a conclusion in section 5. 

The contribution of this paper is to asses if assets in a combined portfolio of wind and- hydropower only should rely on the market when clearing individual imbalances, or if there is potential value obtained by internal coordination in addition to interaction with an intraday pay-as-bid market. The topic is evaluated qualitatively with focus on some identified market and operational challenges as well as quantitatively with case studies for an actor managing wind- and hydro assets.

\section{Problem description}


The topic of wind balancing using hydropower has been discussed in several papers.
The theme has been evaluated from a system point of view where topics related to how large scale penetration of wind in Europe can be balanced by hydropower \cite{OlssonMagnus2009Oohb,doi:10.1260/0309-524X.37.1.79, 2cf18bec461945b8beb9edac8d04dc10,9221980}, but has also been addressed from a portfolio perspective in \cite{8469857}. Aspects related to bidding combined hydro-generation and wind energy to the spot markets have been addressed in \cite{7438907}. In \cite{ANGARITA2007393, 1318699} optimal bidding- and operation strategies of wind-hydropower are suggested. Mathematical formulations are presented for hourly optimization incorporating the stochastic characteristics for wind power. 

The additional contribution in this article is related to value creation obtained by internal handling of complementary imbalances in a pay-as-bid market with shifting liquidity and bid-ask spreads. It is primarily addressing the real-time hourly optimization with less uncertainty for wind and hydro production, but at the same time with more dynamic and comprehensive marginal cost description for hydropower than presented in previous work. The proposed method with a common load commitment for a portfolio ensures optimal allocation of resources and is suited for real-time automatized load distribution within a portfolio.      

\subsection{Market}
The market framework for this article is a day-ahead centrally cleared auction where bids are made once a day,  followed by intraday and balancing markets where imbalance are cleared, and/or power producers seek to create additional profits. Focus has been on the Nordic market, but resembles that of other liberalized power markets. We assume that power producers in this study act as price-takers when submitting bids.

\subsubsection{Pricing power to balancing and intraday markets}
Power producers in liberalized energy markets have a long tradition for pricing according to marginal cost (opportunity cost) in the spot and balancing markets. This follows the rationale that price-takers will maximize their profits by bidding their marginal cost \cite{6607389}.  These markets are defined
as uniform or pay-as-clear auctions \cite{1397542} 
They have been successfully applied in the Nordic market for decades together with strict rules against market manipulation and market surveillance from governmental institutions. With the introduction of the Nordic intraday market around at the end of the millennium, power producers where gradually given the possibility to trade in a pay-as bid market.  
This market has been running in parallel with the balancing market, with traded volumes rapidly increasing the last few years (13 TWh 2018, 20 TWh 2019) \cite{Nordpool}. Hydropower producers have been able to choose whether to clear imbalances and trade available volumes in the intraday or balancing market, and as such pricing signals to the two markets have been equal since any arbitrage between the markets would have been captured by market actors. Interactions and trade-offs between participation in the two markets have been investigated in
\cite{8469997}.


New requirements enforcing a stricter balancing responsibility can gradually force power companies and retailers to secure balance of the portfolio before entering into the hour or minute of operation. This might create an clearer distinction between the two markets resulting in different pricing regimes.  One given by the value of securing balance or income early in the intraday market
, opposed to a value associated with real time balancing in the balancing market.

\subsubsection{Market liquidity and large spreads}
Nordpool defines a well-functioning and competitive day-ahead power market as a market
where electricity is produced at the lowest possible price for every hour of the day \cite{Nordpool}.  The Nordpool day-ahead market is divided into several price areas, but liquidity in the market is still sufficient to ensure market clearing at all times.  This is however not the case for the intraday market. This is a pay-as-bid market, and for several of the the price-areas, the liquidity had traditionally been low. This can either be caused by transmission constraints, volumes allocated to other markets, or simply because markets actors do not consider the value of providing volumes to the intraday market sufficiently high for participation. As a result of limited liquidity, the bid-ask spread
have in many cases historically been high. This is a general trend that can be observed for markets with low liquidity \cite{SarkissianJack2016Svav}. The result is that there often is a large gap between the prices that someone is willing to buy for (BID), and the prices the seller is demanding (ASK). 
The large gaps might alone discourage participation from actors used to a pricing regime based on marginal cost(MC). As long as there exist an alternative balancing market based on MC, this has to a large extent been preferred, especially by Norwegian hydropower producers.  For companies with a tradition in pricing according to marginal cost, and strict rules against market manipulation, it might be more challenging to adjust to a market with a new pricing mechanism exposed to for instant algorithmic based bidding \cite{Edoli2016}. 

\subsubsection{Pricing of bids in a pay-as-bid market versus marginal cost markets} Market structures and auctions in the electricity markets is a widely discussed topic in energy politics and research \cite{KAHN200170}. It is not within the scope of this article to elaborate on the pros and cons with choice in relation to different market mechanisms. The objective is to illustrate how introduction of a  pay-as-bid intraday market linked to a uniform price market (spot and balance) might create challenges and opportunities for a power company with imbalances, and how this might effect the choice of whether to clear these imbalances in a market or through internal balancing.  An important characteristic of a pay-as-bid market is that volumes are cleared directly between two market actors. These actors can be located in a different price areas as long as there is available transmission capacity. As an example, a Norwegian hydropower producer can trade volumes directly with a German wind producer. 



\subsection{Wind- and hydro operations}




\subsubsection{Managing imbalances in combined portfolios of wind and hydro}

To evaluate the difference between internal or purely marked based   
balancing for a single operator in intraday market, it is first necessary to describe the cost associated with imbalances for wind- and hydro operations.


\paragraph{Imbalance cost}

Imbalance cost for wind normally reflect the cash flows from 
the clearing of imbalances. If one assumes that the forecasted wind production is traded on the day-ahead market, \cite{5558699} have shown that the average imbalance cost can be calculated according to eq. \ref{eq:imbalance cost} where $\overline{c}_{imb}$ is the average specific imbalance cost per MWh wind power produced in the considered time period, $Q_{act}(t)$, $Q_{forc}(t)$ are actual and forecasted wind power generation in the settlement period, while $t$ and $\pi_{imb}(t)$ is the imbalance clearing price in $t$. 


\begin{equation}
\overline{c}_{imb}=-{{\sum\limits_{t=1}^{N}}(Q_{act}(t)-Q_{forc}(t))\cdot\pi_{imb}(t)\over{\sum\limits_{t=1}^{N}}Q_{act}(t)}
\label{eq:imbalance cost}
\end{equation}

While eq. \ref{eq:imbalance cost} can be applied directly in a one-price clearing system, for a two-price clearing system one has to take into account the fact that the imbalance price depends on the direction of the 
imbalance: where $\pi_{imb,SB}(t)$ is the system buy price and $\pi_{imb,SS}(t)$ is the system sell price in settlement period $t$.

\begin{equation}
\begin{multlined}
\pi_{imb}(t)= \pi_{imb,SB}(t) if Q_{act}(t)<Q_{forc}(t)  \\ 
\pi_{imb}(t)=\pi_{imb,SS}(t)  if Q_{act}(t)>Q_{forc}(t) 
\label{eq:two price}
\end{multlined}
\end{equation}

The method for calculating imbalances can not be seen isolated from the revenues that are generated in the spot market. Even though the calculation in eq. \ref{eq:imbalance cost} generally give a good estimate for imbalance cost when considered over a long time period and assuming that accumulated sum of imbalances are zero, it can give a misleading signal when considering imbalances for a shorter time horizon and specifically hour by hour. If for instant the actual production is higher than the forecast for the majority of hours considered, the imbalance would actually contribute to revenues rather than cost. This does not reflect the actual cash flows that are involved.

A better performance measure for evaluating the real cost for a shorter time horizon is to apply the concept of "cost of imperfect forecast" which also is proposed in \cite{5558699}. Assuming that the optimal revenue would be obtained by bidding the actual production to spot, the hourly cost compared to forecast can be calculated by eq. \ref{eq:cost of imperfect forecast}.

\begin{equation}
c_{imp}=-{(Q_{act}(t)-Q_{forc}(t))\cdot(\pi_{imb}-\pi_{spot})(t)}
\label{eq:cost of imperfect forecast}
\end{equation}

And the average cost can be calculated by:
\begin{equation}
\overline{c}_{imb}=-{{\sum\limits_{t=1}^{N}}((Q_{act}(t)-Q_{forc}(t))\cdot(\pi_{imb}-\pi_{spot})(t)\over{\sum\limits_{t=1}^{N}}Q_{act}(t)}
\label{eq:average cost of imperfect forecast}
\end{equation}


Imbalances can now be calculated by comparing a scenario where the realised production is sold to spot (best case), against
the actual cost and incomes following from the original sales to the DA market and sale and income from an intraday balancing market. 
Eq. \ref{eq:cost of imperfect forecast} and \ref{eq:average cost of imperfect forecast} will be the performance measures that will be used in the analysis, but since we are investigating cost in a two price clearing system eq. \ref{eq:two price} also applies.

\subsubsection{Wind power}

Transmission System Operators (TSO) and regulators are increasingly imposing stricter responsibilities on the market participants related to handling expected imbalances prior to the hour of operation. The term "balancing responsible" is often used \cite{balance_responsible}, and suppliers of power are obliged to either become balance responsible or enter into an agreement with a participant with balance responsibility.  

\paragraph{Balancing responsibility for wind power}
The power producer can either be the owner of the assets in a combined wind- and hydro portfolio, or they might have taken on a balancing responsibility for parts of the portfolio. By balance responsible, we mean the actor who is responsible for submitting daily production plans and balancing power for predefined groups of power plants to the TSO \cite{NBR}.
In this article, it is not separated between fully owned assets and assets that are commercially operated by a power producer even though there might be reasons given in either contracts and/or legislation that require assets within a portfolio to be managed individually.  An example of the latter could be that a wind farm is owned by two parties, but operated by one of the owners. If there exists an agreement between the owners to share imbalance cost, allocation of internal coordination benefits between the two parties must be taken into consideration.

\begin{figure}[] 
\centering
\includegraphics[scale=0.7]{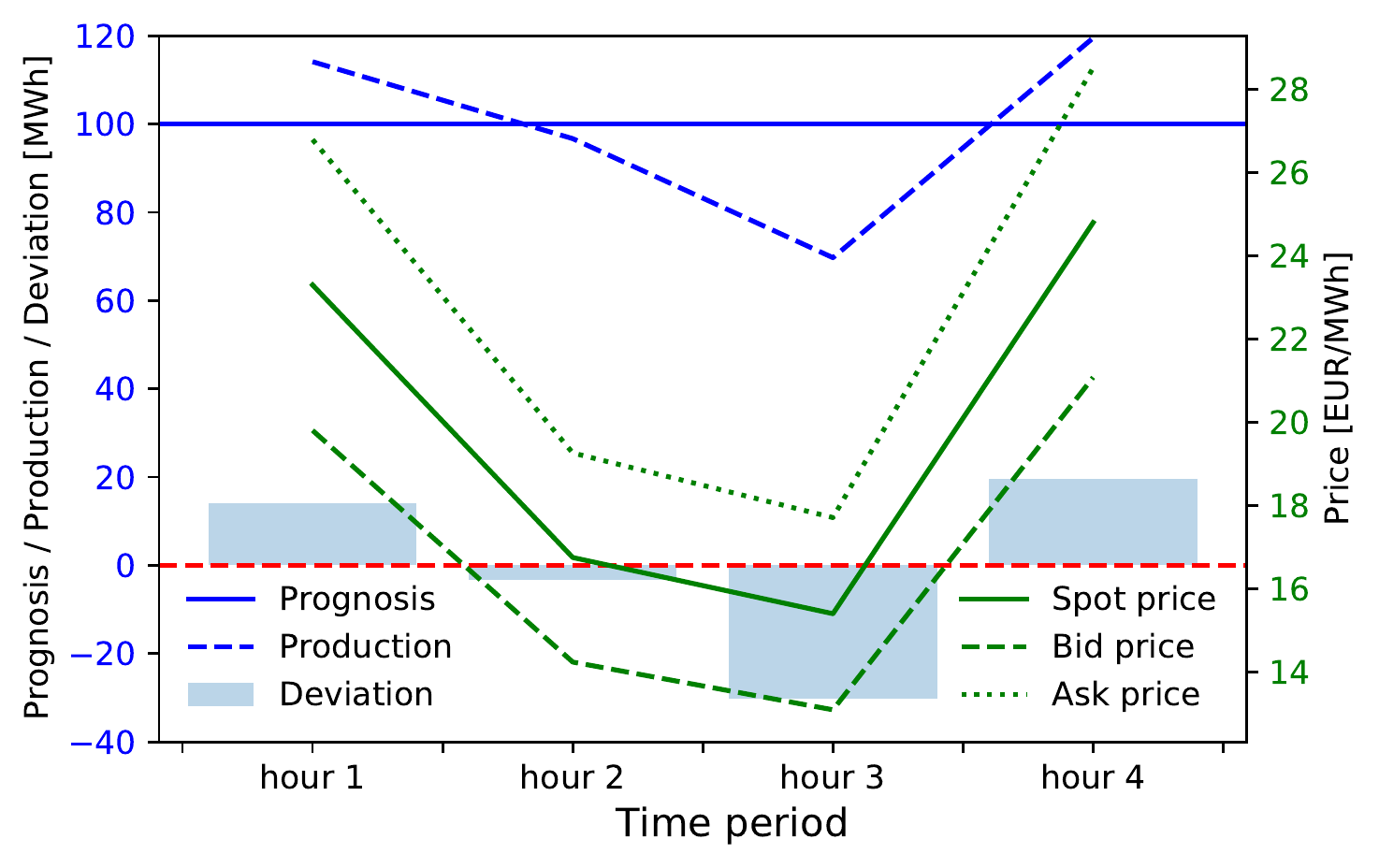}
\caption{Wind production(dashed blue line) versus volume bid to the spot market (blue line). Spot-, bid- and ask prices (green lines) and difference between sold-  and actual production (blue bar) } 
\label{fig:wind_4_seg}
\end{figure}

\paragraph{Imbalance cost for wind}
Fig. \ref{fig:wind_4_seg} represents 4 hours in a day where the power producer has sold 100 MW for all hours at spot prices indicated by the blue line. The actual wind production is according to the dashed blue line. The intraday prices are required to calculate the imbalance cost, and the bid-ask prices in this simplified example is assumed to be -+15\% of the spot price.  Eq.\ref{eq:average cost of imperfect forecast} can further be used to calculate the average imbalance cost to 0.5 EUR/MWh.

\subsubsection{Hydropower}
Calculations related to imbalances cost for hydropower resembles those used for wind power, especially when it is related to run-of river hydro. For reservoir hydro there are some clear distinctions. 
The first relates to the possibility to store water, introducing regulating capacity but also an additional complexity related to valuation of the stored energy, e.g. using the water value method \cite{4501045,WOLFGANG20091642} 


When bidding hydro production into a market, which could be either a day-ahead or intraday market, an established economic principle is to bid the marginal cost of your production. 
Marginal cost for hydropower plant can be calculated by short-term optimisation models using successive linear optimisation (LP) such as in the commercial software SHOP \cite{FossoO.B2004Shsi, AASGARD2016181}. The marginal cost is then represented by the opportunity cost extracted from the LP-problem. Marginal cost can also be calculated from use of heuristics as presented in \cite{7981875}.


Hydropower plants will in most cases have the possibility to regulate discharge through the power plant,  and  there could be significant variation in efficiency and marginal cost associated with different level of operation. 



\begin{figure}[htbp]
\centering
\centerline{\includegraphics[scale=0.4]{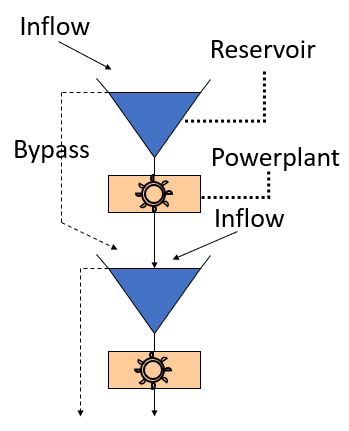}}
\caption{Simple system with two linked power plants. The two illustrated plants can have different marginal cost of production, but how much they influence each other depend on all the factors illustrated in the figure, as well as the discharge capacity for each plant.}
\label{fig:simplesystem}
\end{figure}

For hydrological systems where there are common shared physical constraints between power plants, the value of coordination seems obvious. The simplest case is when an upstream plant produces, and water is lead directly to a downstream plant without reservoir capacity. The downstream plant can then either produce, or let the water by-pass without any income.  
The example illustrated in fig. \ref{fig:simplesystem} is such a system, but in this case with some limited capacity in the downstream reservoir. This cascade will represent the system used further in the analysis for evaluating the benefits of coordination.

\subsubsection{Dynamics for marginal cost (MC) in hydro-based power systems}

\begin{figure}[htbp]
\centering
\centerline{\includegraphics[scale=0.7]{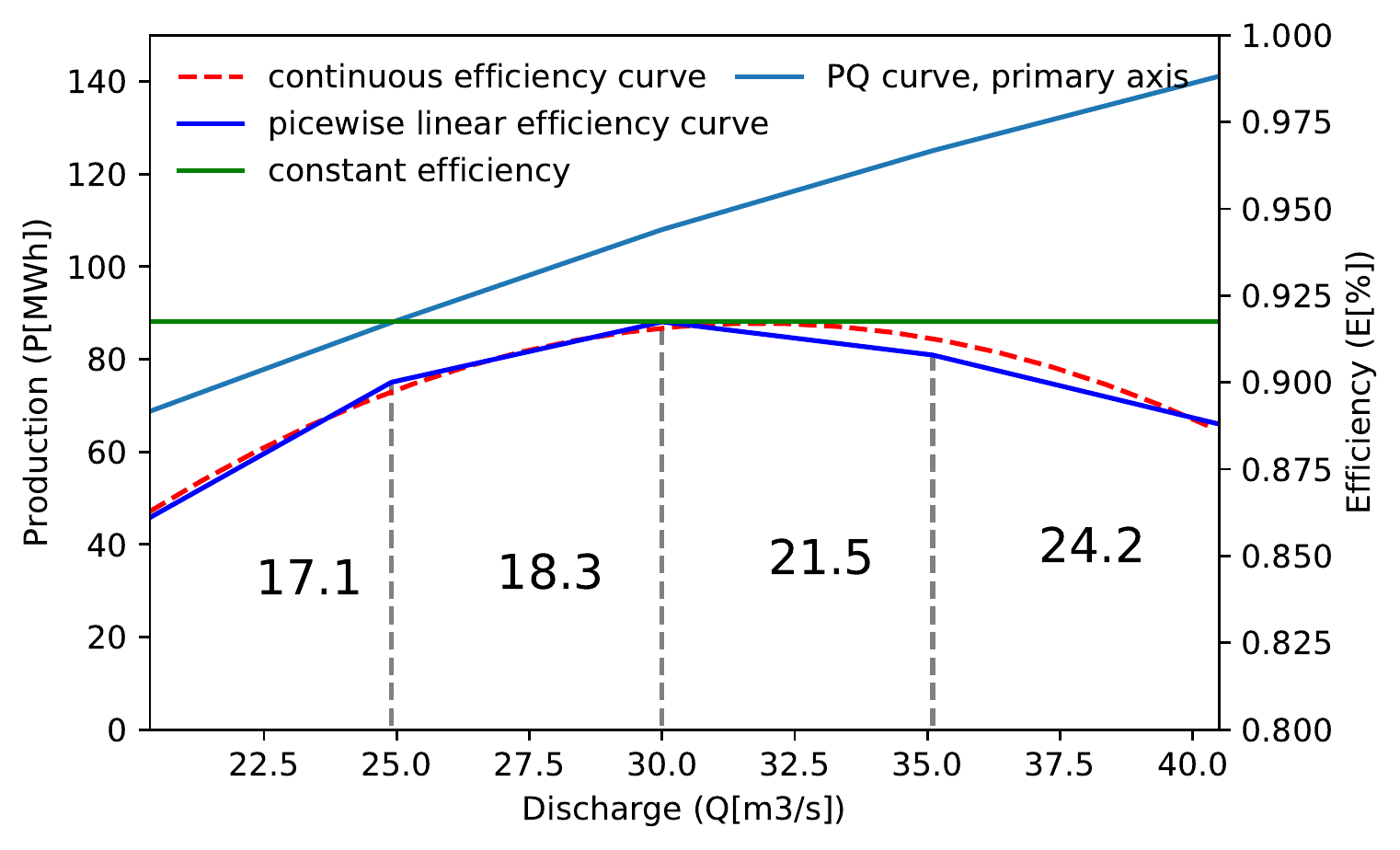}}
\caption{Dynamic marginal cost for a plant with capacity of 140 MW and 42 m3/s plotted in the working range from 68 to 140 MW. The light blue curve illustrates the relationship between plant production and discharge, often referred to as the PQ-curve. The other curves show ways to represent plant efficiency. Marginal cost for the piece-wise linear representation of the efficiency is also shown with numbers} \label{fig:dynamicMC}
\end{figure}

\paragraph{MC only valid for production changes within the same “segment"} In fig. \ref{fig:dynamicMC} different representation of plant efficiency is illustrated. The green line illustrates a plant with constant efficiency. In this case  the marginal cost will also be constant for all levels of production. 
The blue line is illustrating a piece-wise linear efficiency curve. In this case there will be one MC valid for production changes within the discharge segment from 20-25 m/s and another MC from 25-30 m/s and so on. \clearpage
Marginal cost for the piece-wise linear curve can be calculated according to \cite{10.1007/978-3-030-03311-8_7}. Finally, the red dotted line illustrate a continuous shift in efficiency. This is in most cases the most correct representation of efficiency variation in a hydropower plant. This means that there is a marginal cost associated with every point of operation on the production curve. 
Prices to intraday market must be provided as price-volume bids, and to the extreme, every production change has a different price. 

\begin{figure}[htbp]
\centering
\centerline{\includegraphics[scale=0.45]{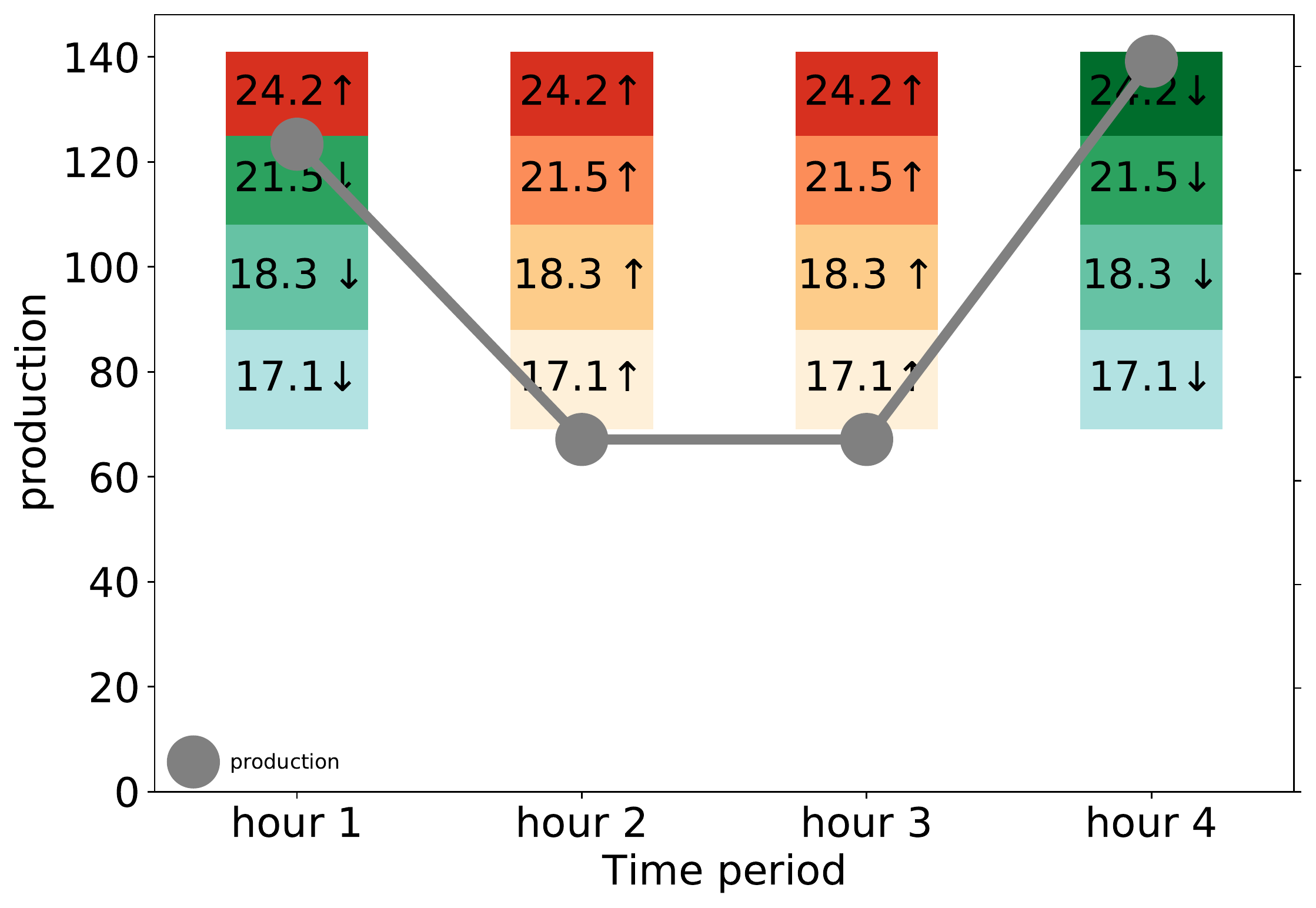}}
\caption{MC breakdown representing how a hydropower producer is providing volumes to the intraday market.  The prices illustrated in fig. \ref{fig:wind_4_seg} are used as input to a river system illustrated in fig. \ref{fig:simplesystem}. The plant efficiency is represented by the piece-wise linear curve illustrated in blue in fig. \ref{fig:dynamicMC}. Green colours indicate bid volumes and prices, while red indicate ask volumes and prices} \label{fig:MC breakdown}
\end{figure}

Fig. \ref{fig:MC breakdown} illustrate how a hydropower producer provide bids to the intraday market. We use the average price for the 4 segments in fig. \ref{fig:wind_4_seg} as water value for the hydropower plant which in this case is 20 EUR/MWh. The  production bid to the day-ahead market 
is indicated by the grey dots in fig. \ref{fig:MC breakdown}. With this production as basis for providing bids to the intraday market, the power producer would be willing to increase production by 16 MWh in hour 1 at a price of 24.2 EUR/MWh. The producer would also be willing to reduce production in this segment. First 17 MWh at 21.5 EUR/MWh, further 20 MWh at 18.3 EUR/MWh, and finally reduce production by 19 MWh to minimum production of 68 MW at a price of 17.1 EUR/MWh. With a piece-wise linear representation of the marginal cost, it makes sense to bid the full segment volume to the market. With a continuous representation of the marginal cost, any point of operation could be selected with a corresponding volume.

\paragraph{Constrained systems might generate “extreme dynamics”}
In cascade hydro systems where production units are placed between large and small reservoirs one can observe   large variations in marginal cost for situations where there is either too much or too little water in the system. If for instance a small reservoir downstream in the cascade suddenly receives more inflow and risk flooding, the marginal cost for the downstream plants might change rapidly from a marginal cost represented by the water value in the upstream plant to zero. Similarly, the upstream plant will receive a clear signal to reduce production to avoid flood downstream dramatically increasing this water-value. 

\begin{figure}[htbp]
\centering
\centerline{\includegraphics[scale=0.7]{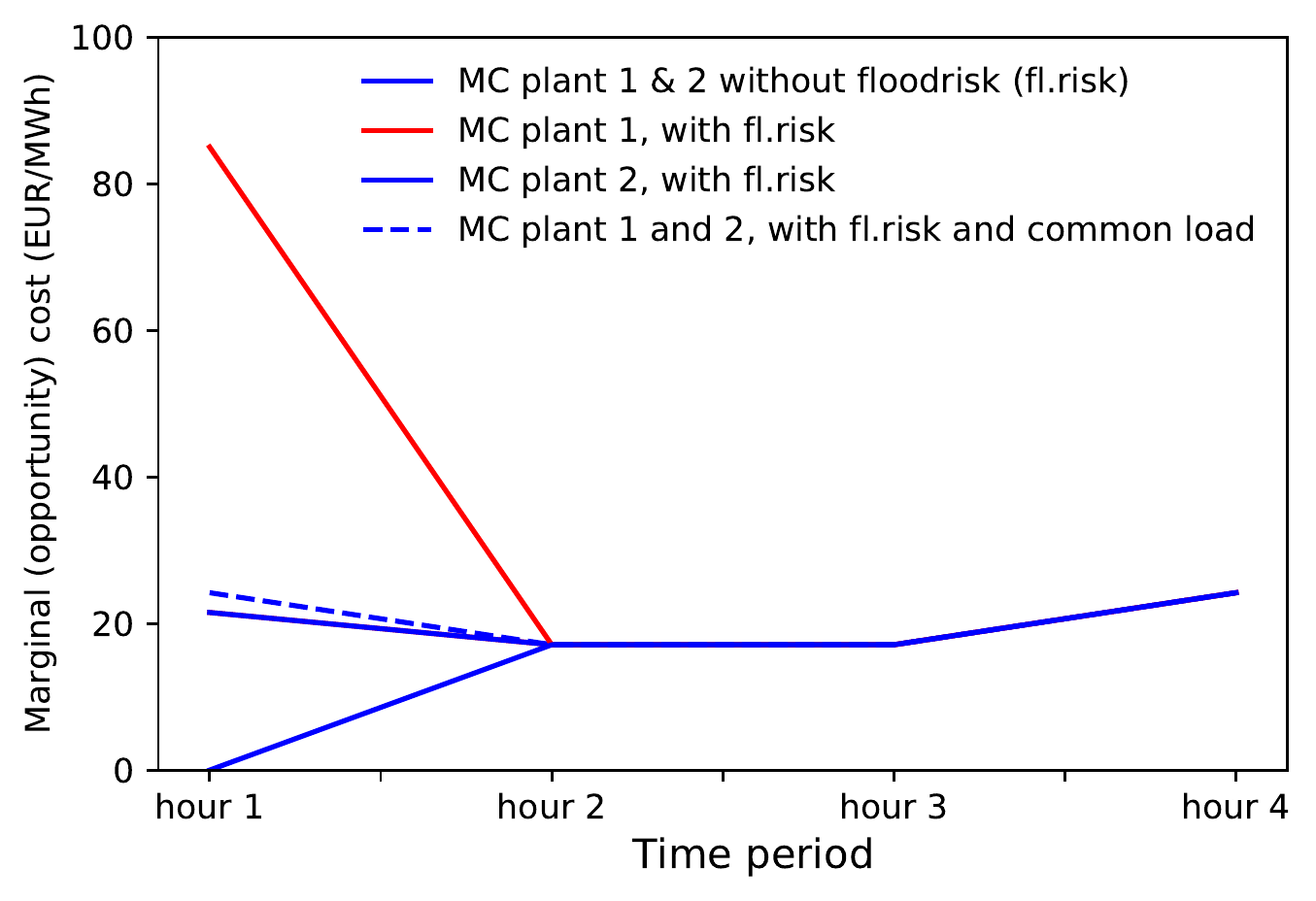}}
\caption{"Extreme" dynamics for MC the cascade river system illustrated in fig. \ref{fig:simplesystem} when exposed to high inflow and risk of flooding} \label{fig:extremeMC}

\end{figure}

Fig. \ref{fig:extremeMC} illustrate an example where inflow to the a downstream plant is delivered above the predictions that formed the basis for the planned production. The upstream plant must either decrease production, or the downstream plant must increase production to avoid flooding. The production and marginal cost in this case are based on the same assumptions as shown in fig. \ref{fig:MC breakdown}. We assume that the efficiency curves and water values are the same for the upstream and downstream plant. We further consider a case where additional inflow is delivered in hour 1 and would lead to flood if production plans remain unchanged. In this case the marginal cost of the downstream plant falls from 21.5 EUR/MWh to zero. This makes sense since the alternative to increasing production is to flood the water. For the upstream plant, the opposite can be observed. The marginal cost jumps from 21.5 EUR/MWh to almost 90 EUR/MWh. The reason for this dramatic increase is that reducing production in this plant, "saves" water for production to a later stage where it can be utilized in both plants. Since the head 
associated with the upstream plant in this example is much lower than for the downstream plant, this effect is reinforced. 


An interesting observation for the example illustrated above is that there in this case are clearly complementary imbalances that can be resolved by internal balancing by moving production intraday for hour 1 from the upstream plant to the downstream plant.  In section \ref{mathrefs} the mathematical formulation associated with moving from plant- to a portfolio load requirement is given. When applying the portfolio load requirement in eq. \ref{eq:loadconstraint_common}, the new marginal (opportunity) cost can be found. The blue dotted line in fig. \ref{fig:extremeMC} illustrate how this can reduce the volatility in MC if the marginal cost are calculated on portfolio level rather than plant level.  



\subsubsection{Pricing of wind imbalances}

For a wind producer, the option of pricing according to marginal cost is less obvious. 
With the very low marginal cost for wind production resulting in wind farms mostly producing at maximum available capacity \cite{5701793}, has shown that it is more relevant to apply opportunity cost based on potential real-time market revenues, or rather the lost opportunity of obtaining these revenues if the capacity is sold or committed to the forward market. 

Even though there exists a wide range of alternatives and strategies related to bidding wind production to the day-ahead market \cite{7078933, VILIM2014141},  production is often provided as a price independent bids to the market based on expected wind forecast. There are national differences on how strict rules are enforced by TSO's 
related to planning production in balance already at the time when bids are placed, but in several markets these requirements will limit the freedom operators have to deviate from the expected forecast when bidding to the day-ahead market \cite{Forskrift}. 

Various approaches have until now been applied to manage the imbalances that occur during intraday operations. These vary from doing nothing, sit back and enjoy life while simply being settled against the balance prices, to more active approaches where intraday market is used to resolve imbalances.  A solution applied by some actors is to only send updated production plans to the TSO based on the expected imbalance production. Within the existing Nordic settlement systems, this would result in limited imbalance cost and potentially imbalance incomes since imbalances contributing in the direction of the system needs will be compensated with profit margins compared to spot \cite{5528939}. 
This settlement system is currently under review, and one expects that there no longer will be potential gains for any plant with deviation from plan during an hour of operation \cite{Singleprice, Guideline}. The requirements related to planning production in balance is also receiving increased attention, and new regulations and incentives will increasingly drive the producers mad. 


However, when imbalance in a wind portfolio occurs, what is the price you should sell or buy the expected imbalance for? The typical approach is to price the imbalance more in accordance with the price observed or expected in the intraday or balancing markets, much in line with the opportunity cost described in \cite{5701793}. Trading algorithms are increasingly being applied to manage bidding which in these cases in larger extent resembles strategic bidding than marginal cost bidding. In \cite{AfsharK2018Obso,5207179} optimal bidding strategies for wind power producers in pay-as-bid electricity market are proposed. Common for the two methods is the use of prediction models for both short term wind power production and intraday prices. Optimisation  could result in a strategy where bidding takes into account the uncertainty of the wind power predictions, which lead to an arbitrage between expected intraday prices and expected imbalance costs.

\subsubsection{Picking or placing bids in the intraday market}

There are two main distinctions when interacting with the intraday market. An actor could either have an infrequent assessment of the market and select/match bids that already are issued. This is typically the case if an actor is in need of resolving an expected imbalance. We can defined this as a reactive approach to intraday market participation. 
Another approach is to actively place bids in the market. This could for instant be done by placing bids that represent available capacity in the portfolio. This approach requires a more continuous follow-up. This is especially important for hydropower producers which in case of a bid published in the market is matched, might need to update the production plan. This might again require that new system information is sent to the TSO, and finally there might be need to re-calculate marginal cost for the remaining portfolio.   


\subsubsection{Time for placement of bids, "first mover" advantage,  and gate closure of markets}
\begin{figure}[htbp]
\centering
\centerline{\includegraphics[scale=0.7]{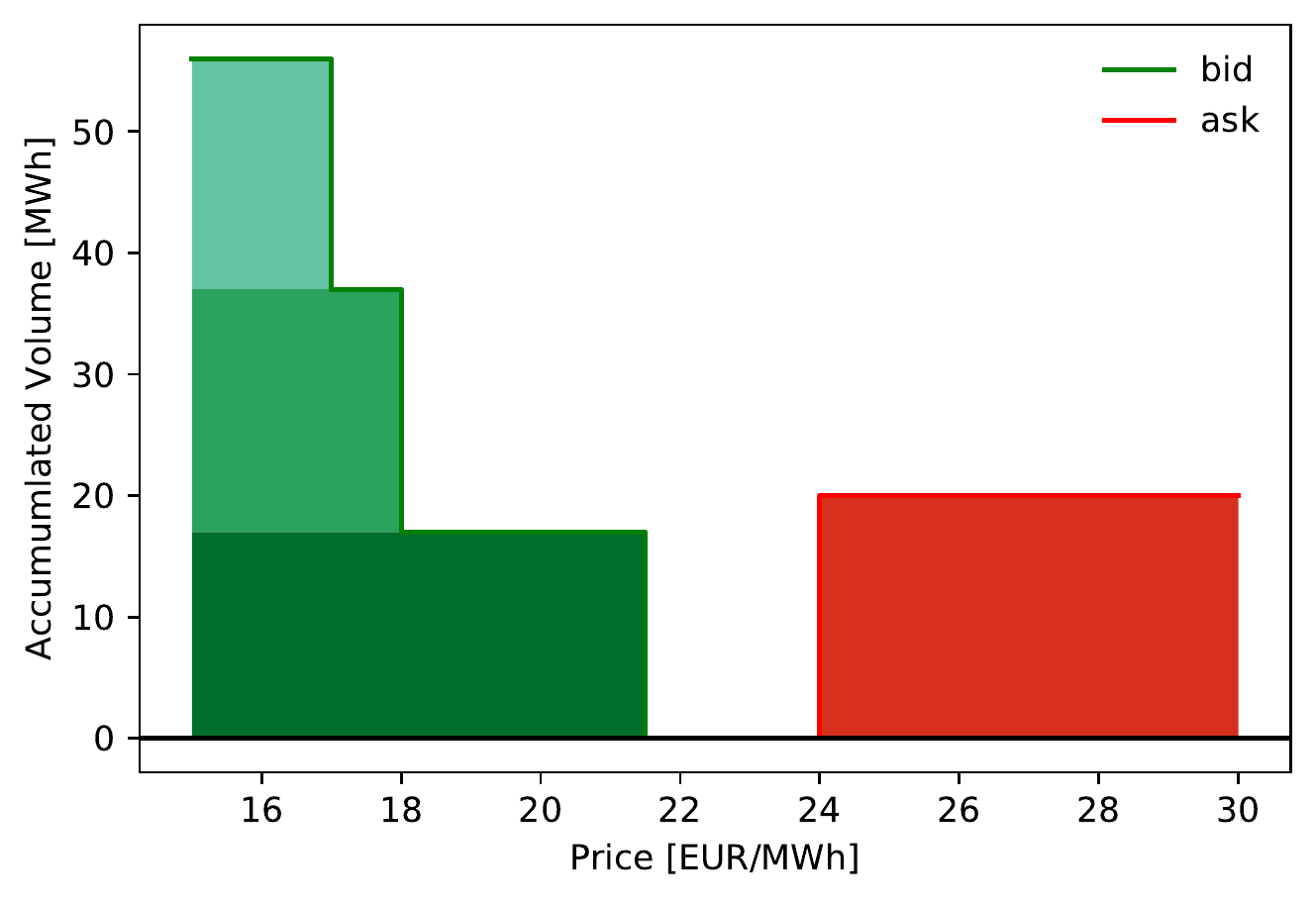}}
\caption{Bid-ask curves. Bid(buyers) side illustrated in green,Ask(sellers) side in red. The bid-ask spread is the gap between the bid and ask side} \label{fig:bidask}
\end{figure}

\paragraph{Time for placement of bids}
The issue related to time for placement of bids can best be described by a simple example. Imagine a company operating two wind power plants (W1 and W2). W1 receives an updated forecast with more wind than expected and need to increase production for the next hour by 15 MWh. Let us further assume that bids to the intraday market are provided by an external hydropower producer with marginal cost indicated in fig. \ref{fig:MC breakdown}. A typical presentation of the bid-ask spreads on an intraday platform is illustrated in fig. \ref{fig:bidask}.

W1 place a sales bid for the next hour at a price lower than the current bid price and will immediately be met by a buy side who will purchase at 21.5 EUR/MWh . Some minutes later, we receive information that there will be too little wind to plant  W2, and production will have to be reduced by 15 MW. We can buy this production from the market at the published asked price at EUR 24.2,  or place a buy bid just slightly above the ask price which would give the same result. Instead of just switching these obligations internally between W1 and W2, we have lost the spread times the volume which in this case is 40.5 EUR. 

If the bids had been placed at the exact same time, this would have placed W1 and W2 as the best seller and buyer in the intraday market and the volumes would be  traded between the two plants. This would however require strict coordination on when bids are placed in the market. 

\paragraph{"First mover" advantage}
Let us assume in this case that W1 and W2 are owned and operated separately, and that both wind operators receive an updated wind forecast at the same time showing that both producers will have imbalances with 15 MWh higher production for the investigated hour the next day. In this case, the first come, first served principle applies in the intraday market, and the first mover is able to clear the imbalance at the lowest cost, represented by the lower green box in fig. \ref{fig:bidask}. Assuming W1 respond first, this plant can sell the excess power at 21.5 EUR/MWh, while W2 must clear the imbalance at the middle green box at 18.3 EUR/MWh. The result using eq. \ref{eq:cost of imperfect forecast} is that W2 end up with higher imbalance cost that W1.
If W2 had been owned and/or operated by the hydro-power producer providing the bids in this example, this could ensure that the first mover advantage is secured internally, and that the externally operated plant would have to clear the imbalance at the unfavorable price. 



\paragraph{Gate closure of markets}
Another important aspect related to placement of bids are closing times for markets. Various gate closure times for the intraday market are being applied throughout national markets, typically ranging from 30 to 60 minutes before the beginning of physical delivery. If information about potential imbalances for the next hour is received later than this, it is not possible to manage this imbalance in the intraday market. The alternative to internal balancing is to enter into the hour with expected imbalances which then will be cleared towards the balancing market.

\subsubsection{Trading and imbalance fees}
Fees for participation in intraday are typically in the range 0.1-0.2 EUR/MWh. (Nordpool, EURONEXT). In addition, producers might  pay fees to suppliers of trading software based on traded volumes in the market. If margins obtained in the markets are put under pressure, the fees associated with trading in the intraday market might be more significant, encouraging increased use of internal balancing.
Power producers also have agreement with TSO's or companies providing settlement services. These might charge fees in connection with imbalances which typically can be in the range 0.15-0.5 EUR/MWh (eSett). This is favouring a strategy where imbalances are solved prior to the hour of operation if cost associated with clearing the imbalance in the intraday or balancing market otherwise are equal.   

\subsubsection{Autonomous trading systems}
Increasing digitization and integration towards market platforms might enable producers and consumers to reflect their true MC in their bids and 
submit these in real time to the market in a larger extent than today. Arguments for internal balancing based on in-house knowledge about physical status will be less prominent, and producers might experience that the benefits of internal balancing might be obtained anyhow since their own information is mirrored in the marketplace.  

On the other hand, with an increasing implementation of autonomous trading systems we will most likely see an increase in the application of trading robots, algorithms and strategic bidding in the market. This could support the use of internal balancing where the real time true marginal cost in the hydro portfolio can be used for internal balancing without exposure to a more unclear market representation.

Finally, the use of autonomous trading systems will require extensive monitoring  and quality assurance of input- and output data. The market actor will still be responsible for all interaction with the market, and ensuring that all rules and regulations are followed. The power sector being defined as critical related to security of supply and national security issues, also have limitations related to how tightly integrated market and supervisory control and data acquisition (SCADA) systems can be.



\section{Proposed solution}
The previous section has illustrated that solely relying on a strategy where external markets are used to manage imbalances within a power portfolio could lead to sub-optimal solutions for a power producer. The price signals to and from the market might simply not sufficiently represent the marginal cost that would give the most profitable outcome for the producer. 
A better solution might be a strategy where the true marginal cost generated by internal models are applied and complimented with opportunities that exists in an external market. 

\begin{figure}[h] 
\centering
\includegraphics[scale=0.5]{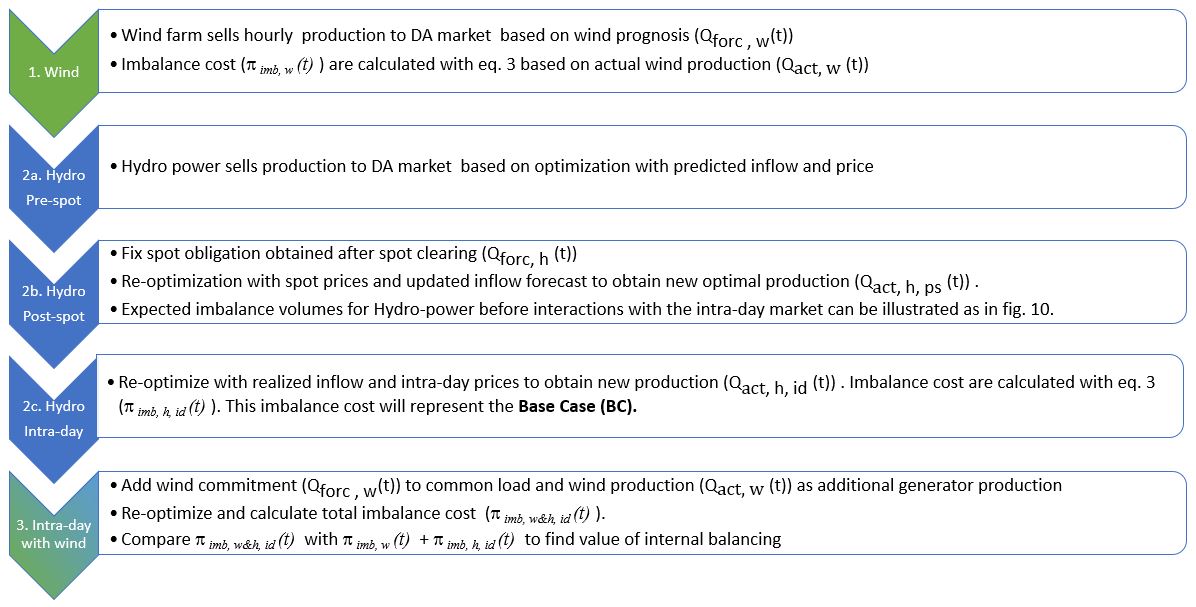}
\caption{A step-wise process and model to evaluate performance of internal imbalance handling compared to individual balancing} \label{fig:process}
\end{figure}

With this in mind, it is interesting to look further into methods for internal balancing. The process for evaluating  the value of internal balancing opposed to individual balancing of each energy resource is illustrated in fig. \ref{fig:process}. 

\subsection{Mathematical formulation} \label{mathrefs}

The processes in fig. \ref{fig:process} starts with tasks that are performed in connection with day ahead bidding. 
Assuming that the wind producer is risk neutral, the producer would bid the expected production to the day-ahead market. 
The optimization problem for a reservoir hydropower producer in a system with predicted market prices can be expressed by the objective function and constraints in eq. \ref{eq:obj}-\ref{eq:seg3}.
The problem is a mixed integer problem and the solution can be found by using Pyomo/Cplex \cite{hart2011pyomo,cplex2009v12}. 


\begin{equation} \label{eq:obj}
Max. \sum_{i}\sum_{t}\left(g_{i,t}*\lambda_{t} \right) + \sum_{m} R_{m}*WV_{m}\\
\end{equation}
s.t.
\begin{flalign}
& Pmin_i \leq g_{i,t}  \leq Pmax_i \hspace{15,5em}     \forall i,t	&\\	
&Rmin_m \leq R_{m,t} \leq Rmax_m \hspace{14,2em}    \forall m,t	&\\
& R_{m,t} = Rinit_m - g_{i,t} + inf_{m,t} - fl_{m,t} \hspace{9,7em}  \forall m, i = 1,  t=1 &\\
& R_{m,t} = Rinit_m + g_{i-1,t} - g_{i,t} + inf_{m,t} - fl_{m,t} \hspace{6,3em}  \forall m,   i > 1,  t=1  &\\
& R_{m,t} = R_{m,t-1} - g_{i,t} + inf_{m,t} - fl_{m,t} \hspace{10em}   \forall m,   i = 1,  t >1 &\\
& R_{m,t} = R_{m,t-1} + g_{i-1,t} - g_{i,t} + in_{m,t} - fl_{m,t} \hspace{7,2em}   \forall m,   i > 1,  t > 1  &\\
&q_{SEG,i,n,t} <= Z_{i,n} \hspace{18,15em}      \forall i,t, n<=1 \label{eq:seg1}	&\\	
\label{eq:seg2}
&q_{SEG,i,n,t} <=  Z_{i,n} * \mu_{i,n-1,t} \hspace{14,4em}      \forall i,t,n>1	&\\
\label{eq:seg3}
&q_{SEG,i,n,t} >= Z_{i,n} * \mu_{i,n,t} \hspace{15,4em}       \forall i,t,n & 
\end{flalign}



The objective function in eq. \ref{eq:obj} optimizes the profits related to selling power to the day-ahead market given by the hourly prediced prices $\lambda_{t}$. Here, $g_{i,t}$ is generation from unit $i$ in time-step $t$, $R_{m}$ is the end reservoir level for reservoir m, and  $WV_{m}$ is the water value for reservoir $m$. 
 
$\mu$ is a binary variable used to control use of water from the different segments. One can not use water from a higher segment before the lower segment is at maximum utilisation. Eq. \ref{eq:seg2} ensures that that segment $n-1$ is "activated" before segment $n$, while eq. \ref{eq:seg3} ensures that all previous segments are maximized before segment $n$ is used. $Z_{i,n}$ is the "capacity" of the segment $n$ for plant $i$ in m3/s. 

The optimisation conducted in process step 2a using eq. \ref{eq:obj}-\ref{eq:seg3} will result in a spot commitment for the hydropower. The next steps are steps associated with the intraday optimisation (step 2c) where trading of imbalances in an intraday market are included in the model formulation.


\begin{equation} \label{eq:maxwload}
Max. \sum_{i}\sum_{t}\left(L_{i,t}*\lambda_{t}-g_{BUY,i,t}*P_{ASK,t} + g_{SELL,i,t}*P_{BID,t} \right) +  
\sum_{m} R_{m}*WV_{m}
\end{equation}

\begin{equation} \label{eq:scheduleconstraint}
g_{i,t} = L_{i,t}-g_{BUY,i,t} + g_{SELL,i,t}
\end{equation}

$P_{BID,t}$ and $P_{ASK,t}$ are intraday bid and ask prices in  time step $t$.
$g_{BUY,i,t}$ and $g_{SELL,i,t}$ are the optimal volumes to be bought and sold in the intraday market in for unit $i$, timestep $t$. 
$L_{i,t}$ is the load requirement unit $i$ in timestep $t$.

For the final step in the process (step 3) we modify the load commitment to represent a common portfolio commitment. The generator production for wind is also added as a source of supply ($g_{wind}$). 
\begin{equation} \label{eq:loadconstraint_common}
g_{i,t} = L_{t} - \sum_{i} g_{BUY,i,t} + \sum_{i} g_{SELL,i,t}
\end{equation}
The constraints in eq. \ref{eq:scheduleconstraint} and eq. \ref{eq:loadconstraint_common} might seem similar , but there is a fundamental difference in the requirement that potentially will have a large impact on the objective function and marginal cost.  Constraint eq. \ref{eq:scheduleconstraint} is 
a plant schedule constraint, and requires that each plant has to meet the specific load requirement that was allocated to this plant during the optimisation towards market prices. 
Constraint eq. \ref{eq:loadconstraint_common}, only requires that the load requirement is met in total, but that this can be met by the combined production from all plants. 

\section{Case study}

\paragraph{Analysis of one day operation of hydro and wind power with exogenous market description}

To investigate the effect of internal balancing for a portfolio consisting of both wind and hydropower, a realistic case 
has been investigated by applying the process described in fig. \ref{fig:process}.

The objective behind applying this case study is primarily to illustrate the concept and interactions in a portfolio with wind and water assets, and not to quantify the long term effects of coordination.  One day is therefore selected to illustrate the concept, a day where there is imbalances both in the wind and hydropower plant. 

Both plants have bid 
their production to the Nordpool day-ahead market. The wind plant has sent bids based on the the forecast (expected) wind prognosis that is available before the bidding deadline (12-noon), this often based on the EC00 \cite{ECMWF} prognosis which typically is available at 8 am. The hydropower has sent bids based on an optimisation with predicted prices and inflow. 

\begin{figure}[htbp]
\centering
\centerline{\includegraphics[scale=0.7]{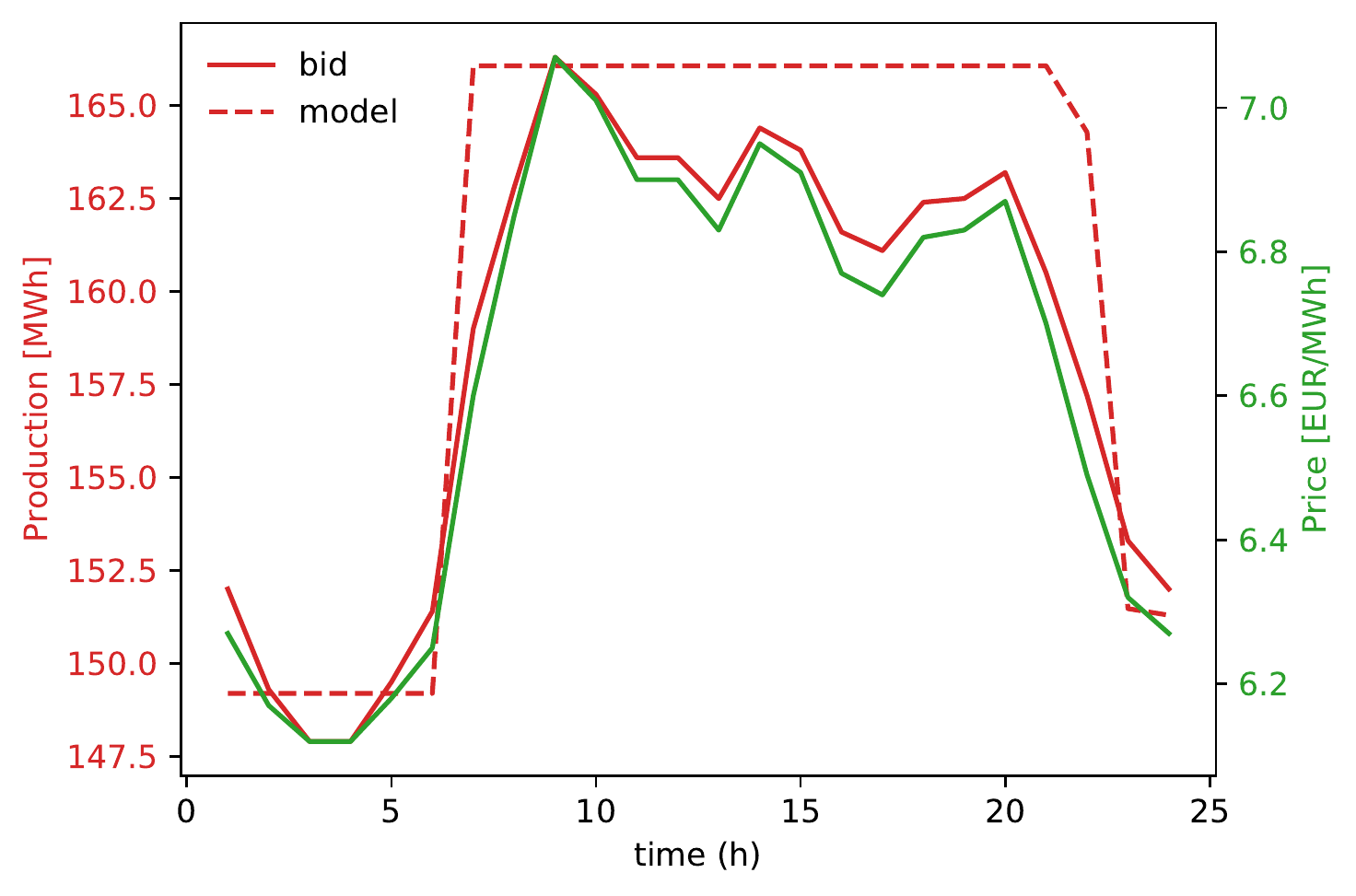}}
\caption{Modelling of next day's hydropower production (Step 2a). "Bid" is the actual production sold to the power exchange (real-life data). "Model" is the approximation made by a simplified optimisation model with  expected inflow and prices for the next day}
\label{fig:model}
\end{figure}

To replicate the bidding process for the hydropower unit, a simple optimisation model has been established based on the equations described in section 3. Given the expected inflow and prices for the next day, the model seems to fit production for the next day well as seen in fig. \ref{fig:model}. The deviation is primarily due to the higher resolution in the description of the efficiency curve in real life operations which will generate a "smoother" production curve for the operational model.

Both plants will after the DA market clearing receive a production commitment for the next day.  
The EC12 model results are also available on a daily basis at approx. 8 pm. 
For this example we assume that the updated results that are available just before entering into the day of operation represent the realised inflow and wind for the next day. No additional uncertainty is considered.

\begin{figure}[htbp]
\centering
\centerline{\includegraphics[scale=0.7]{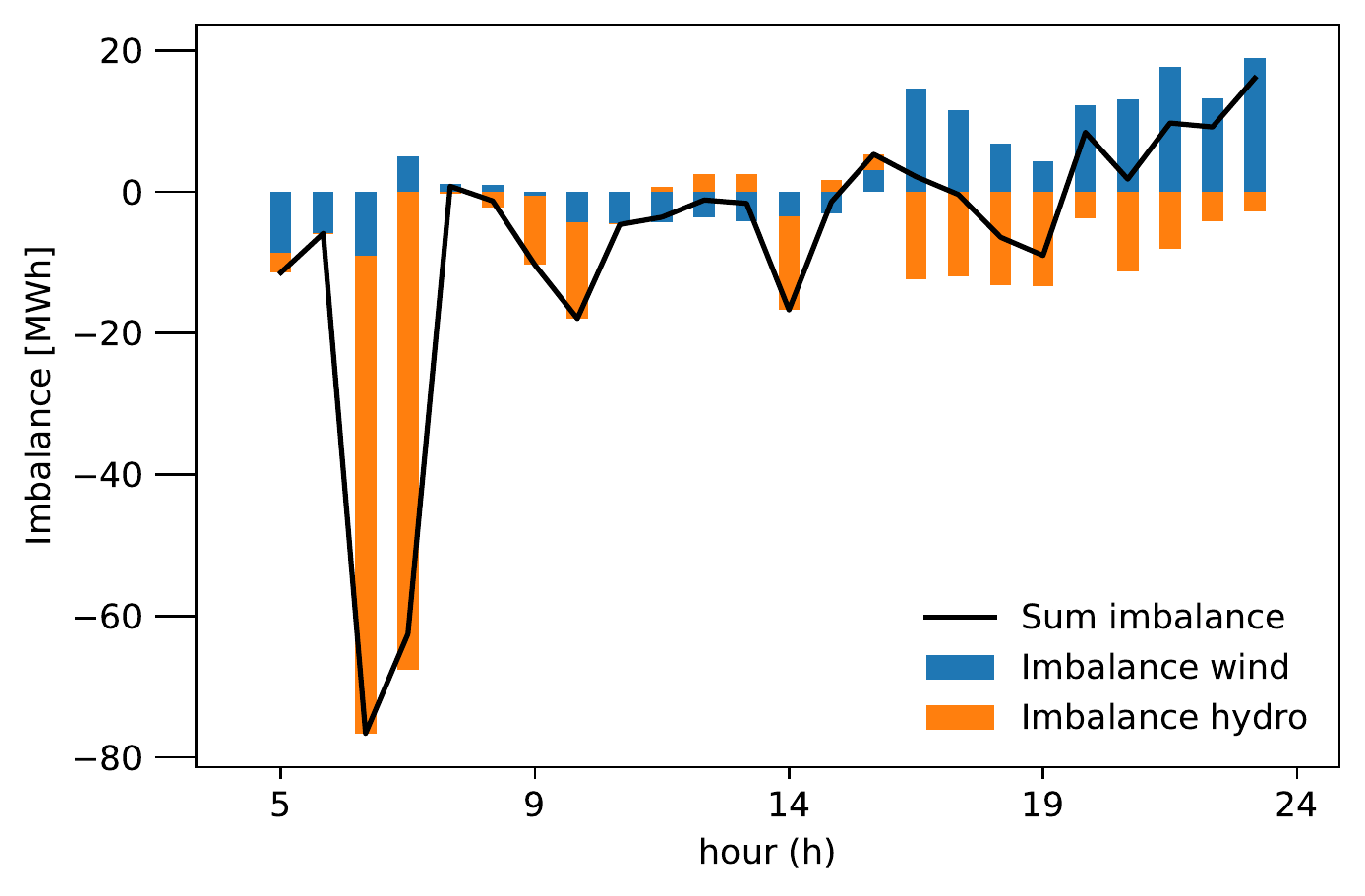}}
\caption{Imbalances for wind- and hydropower as result of updated forecast. Sum imbalance is the net hourly imbalance for a portfolio before re-optimisation towards an intraday market.}
\label{fig:imbalance}
\end{figure}

Based on the updated forecast we find that the expected imbalances are as illustrated in fig. \ref{fig:imbalance}.
The imbalance for wind is simply the difference between the forecast provided at 8 am and 8 pm, while the imbalance for hydro is based on a recalculated plan based on a new inflow forecast. 

Imbalance for unregulated hydropower resembles wind power to a large extent, while imbalance for well regulated hydropower is seldom an issue due to the possibility to adjust for deviations between planned and realised inflow by using the reservoir capacity.  There exist however quite a lot of smaller hydropower reservoirs with downstream plants where there exist some flexibility in short term planning, but where changes in inflow might require adjustments to the existing plan to avoid flooding or running out of water.

\begin{figure}[htbp]
\centering
\centerline{\includegraphics[scale=0.7]{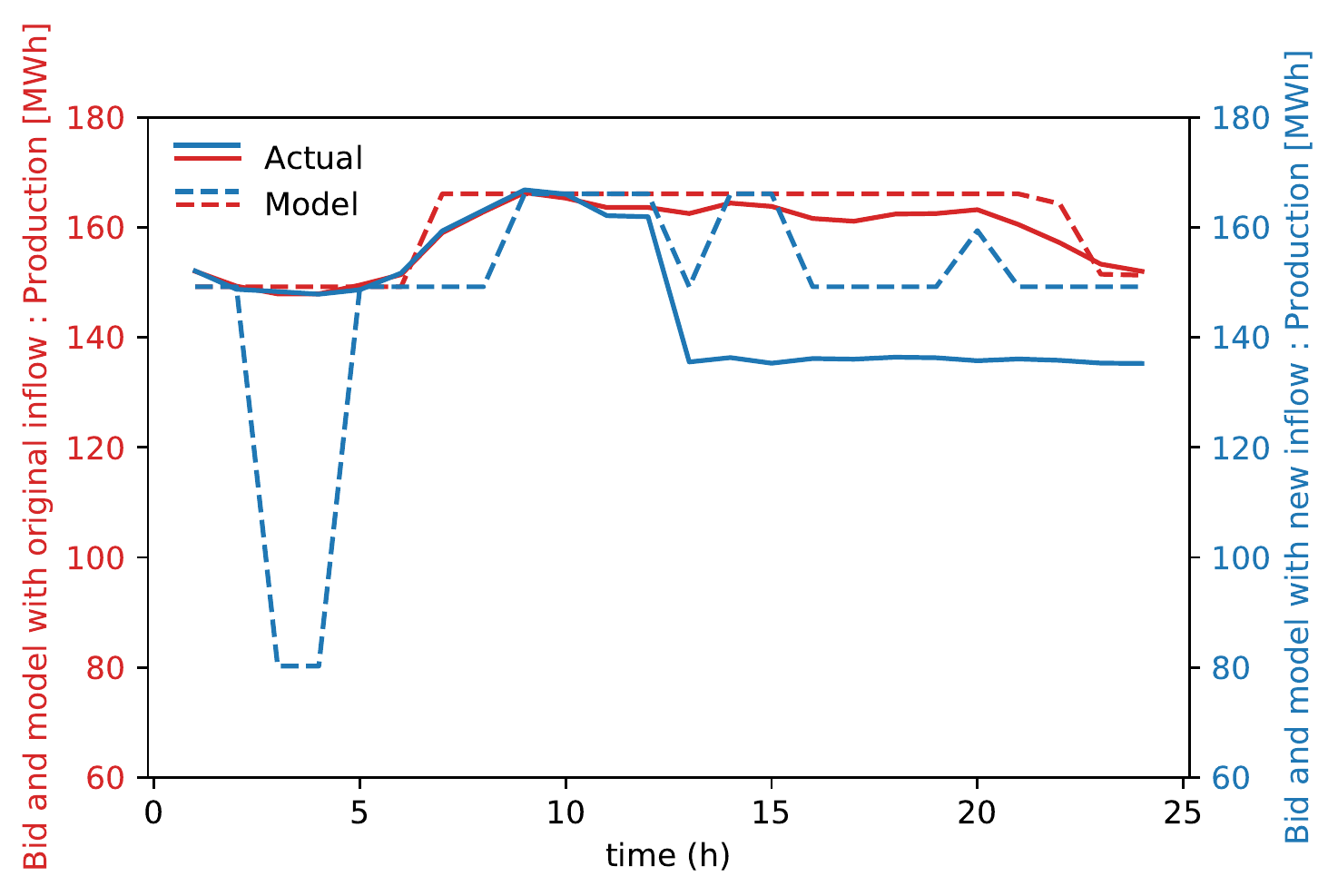}}
\caption{Modelling of next days hydropower production with updated inflow forecast (step 2b). Due to a forecast with less inflow, there is a considerable reduction in modelled production in hour 3 and 4 compared to the optimisation conducted in step 2a.}
\label{fig:model_newinflow}
\end{figure}

The plant and reservoir considered in this case study is such a plant. If no additional market information is provided to the optimisation model, the recalculated plan based on a new inflow forecast for such a reservoir will attempt to maximise the profits based spot prices for the next day and the updated inflow forecast. Fig. \ref{fig:model_newinflow} illustrates how the historic production turned out to be for the investigated plant,  compared to results from the simplified optimisation model. While the historic production follows the original plan until the plant at a certain stage have to reduce production due to lack of water, the optimisation based on a updated forecast seem to more actively exploit the price differences to reduce production in periods with lower prices. The total production over the 24 hours is equal.  
The first approach to evaluate the benefit of coordination is to merge the commitments for the hydro and wind producer. Further, the total cost for the the common realised production cleared against intraday prices is compared to a scenario where the imbalances are cleared individually for wind and hydro.

Imbalance cost are calculated according to eq. \ref{eq:cost of imperfect forecast} were the hourly imbalance cost are summed to give a total imbalance cost for one day. This approach to internal balancing is defined as \textit{reactive} since we are not conducting any re-optimisation for the common load, but only exploiting the value of any complimentary imbalances. Results from such a reactive approach to coordination is illustrated together with other approaches in table \ref{table:results}.

\paragraph{Introducing a market}

\begin{figure}[htbp]
\centering
\centerline{\includegraphics[scale=0.6]{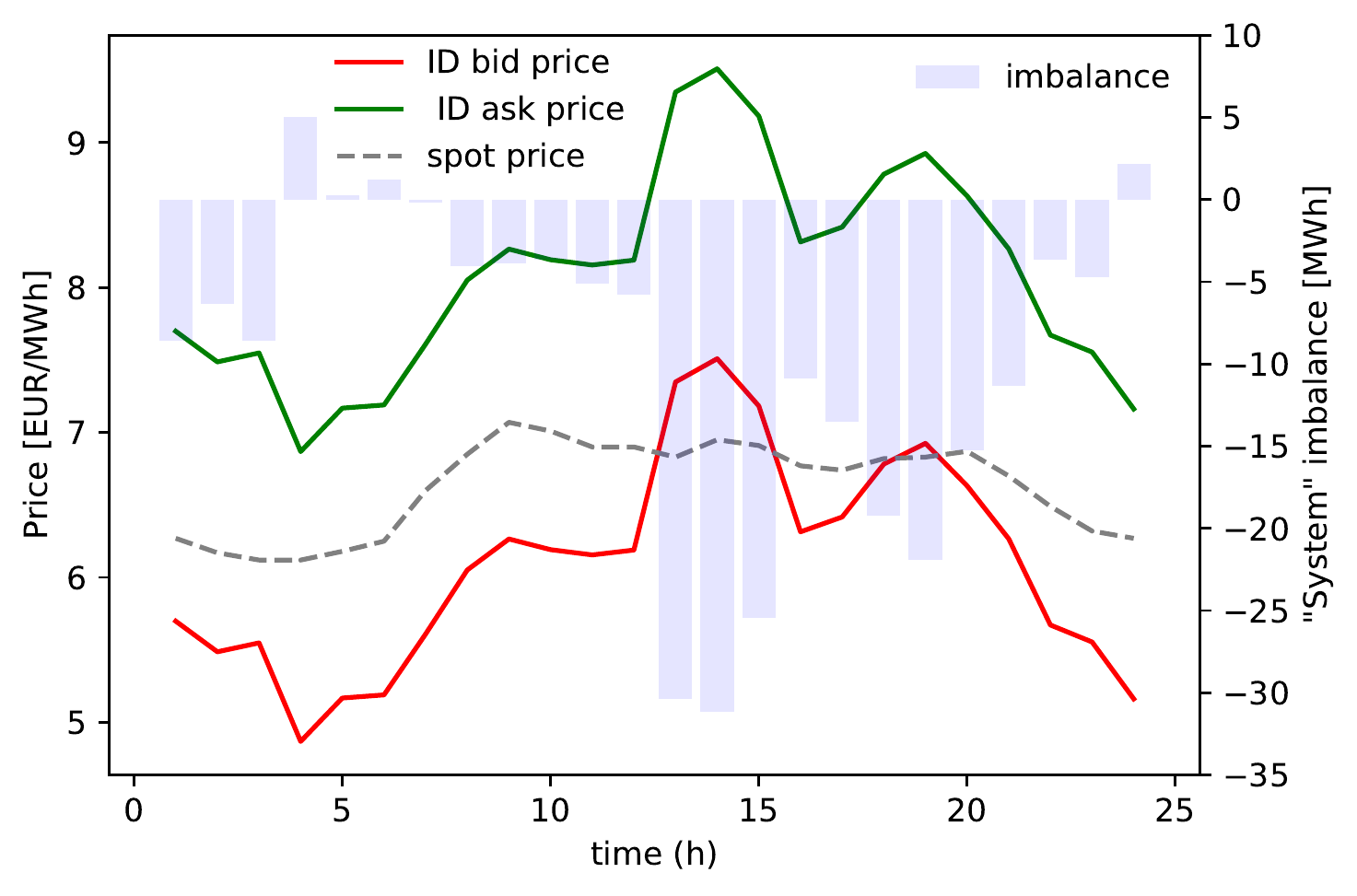}}
\caption{Synthetic intraday prices for Oct 7. 2020, with the realized NO2 spot price for the same day. Bid- ask prices are calculated with fixed margin of +-15 \% of the spot price and a correction factor of +- 1 \% for each MW of imbalance between E00 and E12 prognosis illustrated by the bars in the chart. The "system"-imbalance in this graph should not be confused with the imbalance in fig. \ref{fig:imbalance} which is the "producer"-imbalance after re-optimisation of the hydro-power production}
\label{fig:syntetisk}
\end{figure}

To be able to calculate imbalance cost, a market description for intraday trading is required.
The historical bid-ask prices related to the intraday market are not easily accessible. These are dynamic prices that change continuously, and getting a "snap-shot" of the market for instance just before entering into a new day requires access to order books containing large amounts of information. Order books can typically be provided by the market operators at some fee.  High-, low, and last prices are more accessible, but these tend to deviate considerably from observations made in the market several hour ahead of the closing time for each hour. 
To represent the market conditions in our case study, a synthetic market description has been generated. We assume the the bid and ask spreads are given by a fixed margin of +-15 \% of the spot price. It is also expected that bid-ask prices are effected by the "system" imbalances that are expected the next day. We therefore add a correction factor of +- 1 \% for each MW of imbalance.
This "system"-imbalance is purely calculated as the hourly delta of inflow and wind in MWh between the ECOO and EC12 prognosis. These figures are not calibrated towards market observations, and the sensitivity factor of bid- ask- prices would be very different in a market with considerably more volume. It still illustrates some of the variations that can be expected in the intraday market, and help to illustrate some of the dynamics that arise when bid-ask spreads change throughout the day. Fig. \ref{fig:syntetisk} illustrate the prices used further in the analysis.

\paragraph{Re-optimisation of hydropower}
As soon as a market description for the next day is available, it is possible to actively re-schedule the next day's production. In the Nordic market, the intraday market for the next day opens at 2 PM .
If we assume the the bid- ask prices illustrated in fig. \ref{fig:syntetisk} represent the market at the point of time when an updated wind and inflow forecast is present, the actors can attempt to close their imbalances based on the prevailing prices. 

The wind operator has no possibility to move production from hour to hour, and will have to close the forecasted gaps. The hydropower producer has some reservoir capacity and can move production in a way where imbalances are moved away from the hours with high expected  imbalance cost, to hours where it is possible to buy cheap and sell expensive.
According to step 2c in fig. \ref{fig:process}, the new imbalance cost for hydropower is calculated. This is then defined as our base case. The reason for using the new optimisation as base case for evaluation of the value for internal balancing, is because we wish to evaluate the value of wind-hydro coordination, and not the value of improved optimisation of the hydro-power plant alone. 

The optimised uncoordinated benchmark is the sum of this re-optimised imbalance and wind imbalance and is shown in table \ref{table:results}. The price input in this model is a synthetic and simplified representation of the intraday market. In real life, the bid and ask prices are linked to volumes. Gradually increasing supply/demand in the market is normally associated with gradually lower/higher prices.
One should therefore expect that if the hydropower producer attempts to move production from high-imbalance-cost hours to hours with more favorable prices, the prices would actually respond in a way limiting the value of changing production. This secondary effect is not considered in the presented case study.     

\begin{figure}[htbp]
\centering
\centerline{\includegraphics[scale=0.7]{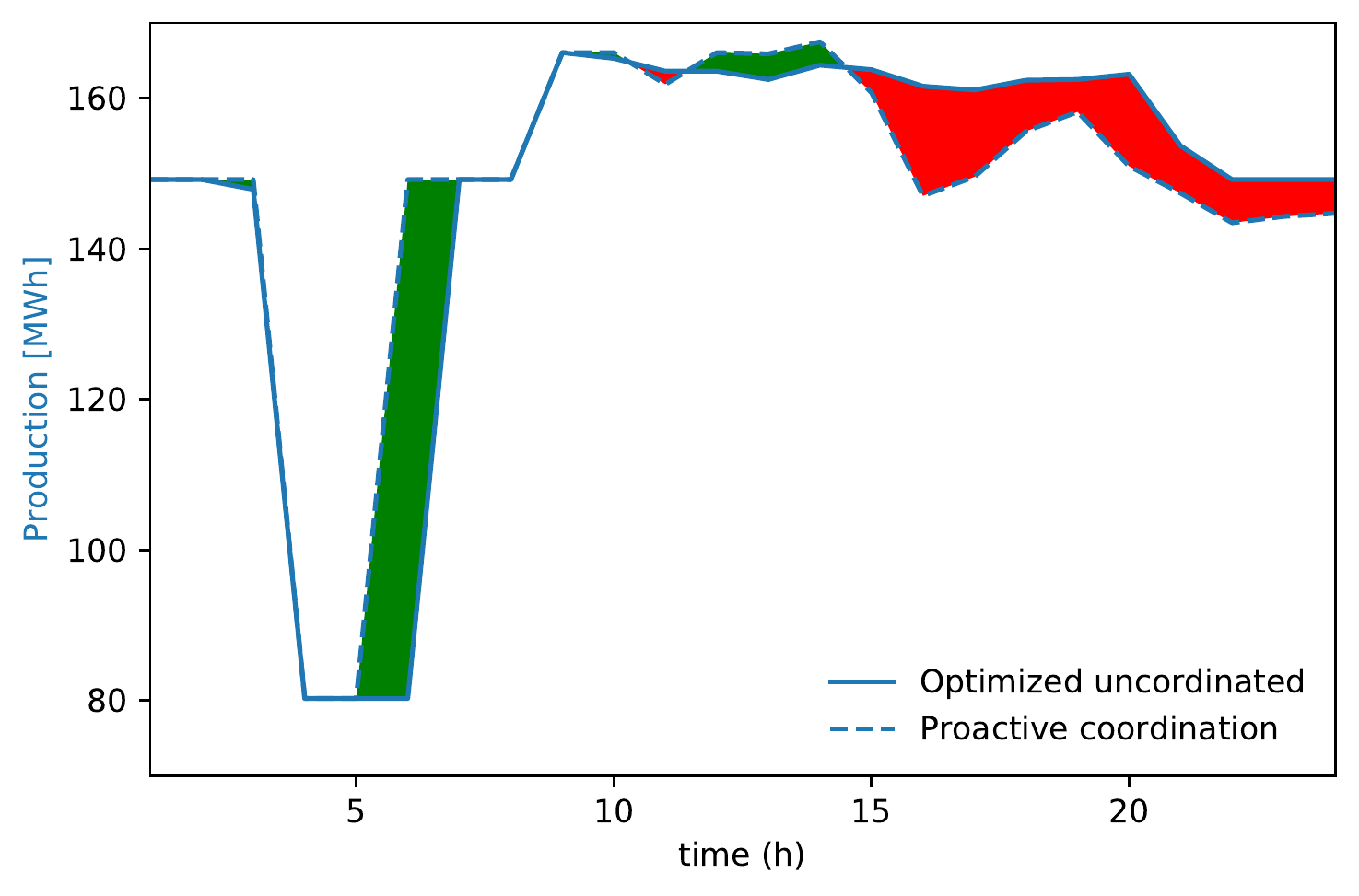}}
\caption{Changes in hydropower production as result of internal balancing. Green illustrate where the hydropower plant increase production compared to base-case, while red illustrate reduced production. The total production is equal and limited by the availability of water in the downstream reservoir.}
\label{fig:proaktiv_opt}
\end{figure}

The final step (step 3 in fig. \ref{fig:process}) is to add the wind production , and optimize towards the common commitment using the load constraint given by eq. \ref{eq:loadconstraint_common}. The results are shown in table \ref{table:results}, and illustrates that the value of internal balancing for this day is 170 EUR, given by the reduced imbalance cost for the coordinated scenario compared to the uncoordinated alternative. 

Comparing with the reactive coordination approach the additional value of proactive balancing is 115 EUR. While the reactive approach is able to create value by exploiting the complementary imbalance in the wind and hydro production, the active re-optimisation creates additional value by using the flexibility in the hydropower reservoir. Fig. \ref{fig:proaktiv_opt} illustrate how hydro production is decreased in periods where wind compliment the hydro imbalance, and that the production can increase in hour 6 where there is a considerable imbalance cost for the hydropower plant. The imbalance cost in hour 4-6 follows from the re-optimisation in the intraday market (step 2c), and the hours are chosen by the model due to the relatively low system-imbalance and favourable ask prices in this period.  


\begin{table*}
\caption{Results from approaches with individual- versus internal balancing. Scenario 2 \\ represents the base case.}\label{tbl1}
\scalebox{0.7}{%
\begin{tabular}{c c c c c}
\toprule
Scenario & Scenario & imbalance cost &  Average IBC and IBC as & Savings compared \\
nr. & name & (IBC)& percent of total income & to base-case \\
\midrule
1&Individual balancing, no optimisation & 500 EUR & 0.13 EUR/MWH, 2.0\% & \\
2&Individual balancing, hydro optimized, base-case & 390 EUR & 0.10 EUR/MWH, 1.5\%   & \\
3&Common commitment, reactive, hydro optimized & 335 EUR & 0.09 EUR/MWH, 1.3\% & 55 EUR \\
4&Common commitment, proactive, all optimized & 220 EUR & 0.06 EUR/MWH, 0.9\% & 170 EUR \\
\bottomrule
\label{table:results}
\end{tabular}}
\end{table*}

\section{Conclusion}
In this article we have shown how the imbalance cost for a power producer with both wind and hydro-power assets can be reduced by internal balancing in combination with sales and purchase in a pay-as-bid intraday market. Knowledge about the marginal cost pricing method for hydropower production is important to understand and optimize the interactions in the balancing process. The potential rapid changes in marginal cost that can be observed for hydropower,  and need to select volume- and price pairs when bidding to a pay-as-bid market, might in some cases favour internal balancing rather than clearing imbalances in the market. 
For a realistic case study from the Norwegian power market, we have demonstrated how a step-wise process can be applied to quantify the value of internal balancing, opposed to an uncoordinated approach from the hydro- and wind production. We have not attempted to answer if the same results could be obtained if the hydropower producer had issued the available balancing power to the intraday market. There are however many aspects in that process that could lead to optimality for the system, but end up as sub-optimal for the producers managing both assets.
Quantifying the long term effects by simulating over a longer time horizon with historic intraday prices could be a topic for further research. 

An important question is what happens if an increasing amount of participants choose to conduct a large share of balancing internally rather then using the market as the primary source for clearing of imbalances. Who will then provide capacity to the intraday market? If liquidity in this market increases which further could lead to decrease in bid-ask spreads, the incentive for internal balancing will be reduced. Power producers should therefore have an incentive to increase liquidity in this market, and internal balancing should therefore be limited to cases where interactions with the market are challenging due to time restrictions or dynamics in the system where it is difficult to publish and follow-up true marginal cost in the market.

\bibliographystyle{elsarticle-num}
\bibliography{cas-refs}

\end{document}